\documentclass{article}
\usepackage[T1]{fontenc} 
\usepackage[utf8]{inputenc} 
\usepackage{ismir,amsmath,cite,url}
\usepackage{graphicx}
\usepackage{color}
\usepackage{todo}
\usepackage{cite}
\usepackage{amsmath,amssymb,amsfonts}
\usepackage{inconsolata}
\usepackage{algorithmic}
\usepackage{graphicx}
\usepackage{multirow}
\usepackage[hidelinks,colorlinks=true]{hyperref}
\usepackage{url}
\usepackage{textcomp}
\usepackage{xcolor}
\usepackage{tabularx}
\usepackage{booktabs}
\usepackage[noindentafter]{titlesec}
\usepackage[caption=false,font=normalsize,labelfont=sf,textfont=sf]{subfig}
\usepackage[noabbrev,capitalise]{cleveref}
\usepackage{enumitem}
\setlist{parsep=0ex,topsep=0.5ex,itemsep=0ex,leftmargin=2em}
\definecolor{myblue}{rgb}{0,0.3,0.6}
\hypersetup{linkcolor=myblue,urlcolor=myblue,citecolor=myblue,anchorcolor=myblue}

\newcommand{\tba}[1][]{\textcolor{blue}{TBA\ifx&#1&\else---#1\fi}}

\definecolor{darkblue}{rgb}{0.0, 0.0, 0.55}
\definecolor{darkred}{rgb}{0.55, 0.0, 0.0}
\definecolor{darkgreen}{rgb}{0.0, 0.2, 0.13}
\definecolor{darkgoldenrod}{rgb}{0.72, 0.53, 0.04}
\newcommand{\yes}{\textcolor{darkgreen}{\checkmark}}
\newcommand{\no}{\textcolor{darkred}{$\times$}}

\newcommand{\myfnsymbol}[1]{\ensuremath{\ifcase#1 \or * \or \dagger \or \ddagger \or \mathsection \or \mathparagraph \else \@ctrerr \fi}}
\newcommand{\tabfn}[1][]{{\ifx&#1&\textsuperscript{$*$}\else\textsuperscript{\myfnsymbol{#1}}\fi}}
\crefformat{footnote}{#2\footnotemark[#1]#3}
\crefrangeformat{footnote}{#3\footnotemark[#1]#4--#5\footnotemark[#2]#6}
\crefmultiformat{footnote}{#2\footnotemark[#1]#3}{\textsuperscript{,}#2\footnotemark[#1]#3}{\textsuperscript{,}#2\footnotemark[#1]#3}{\textsuperscript{,}#2\footnotemark[#1]#3}
\title{Generating Symbolic Music from Natural Language Prompts using an LLM-Enhanced Dataset}

\multauthor
{
Weihan Xu$^1$ \hspace{1cm} 
Julian McAuley$^2$ \hspace{1cm} 
Taylor Berg-Kirkpatrick$^2$ \hspace{1cm} 
} { 
\bfseries{ \hspace{1cm} Shlomo Dubnov $^2$ \hspace{1cm} Hao-Wen Dong $^3$}\\
 $^1$Duke University\hspace{1cm} 
  $^2$UC San Diego \hspace{1cm} 
  $^3$University of Michigan\\
{\tt\small weihan.xu@duke.edu}
}

\def\authorname{Weihan Xu, Julian McAuley, Taylor Berg-Kirkpatrick, Shlomo Dubnov and Hao-Wen Dong}

\begin{document}

\maketitle

\begin{abstract}
Recent years have seen many audio-domain text-to-music generation models that rely on large amounts of text-audio pairs for training. However, symbolic-domain controllable music generation has lagged behind partly due to the lack of a large-scale symbolic music dataset with extensive metadata and captions. In this work,
we present \textit{MetaScore}, a new dataset consisting of 963K musical scores paired with rich metadata, including free-form user-annotated tags, collected from an online music forum.
To approach text-to-music generation, We employ a pretrained large language model (LLM) to generate pseudo-natural language captions for music from its metadata tags.
With the LLM-enhanced MetaScore, we train a text-conditioned music generation model that learns to generate symbolic music from the pseudo captions, allowing control of instruments, genre, composer, complexity and other free-form music descriptors. In addition, we train a tag-conditioned system that supports a predefined set of tags available in MetaScore. 
Our experimental results show that both the proposed text-to-music and tags-to-music models outperform a baseline text-to-music model in a listening test. While a concurrent work Text2MIDI \cite{bhandari2024text2midigeneratingsymbolicmusic} also supports free-form text input, our models achieve comparable performance. Moreover, the text-to-music system offers a more natural interface than the tags-to-music model, as it allows users to provide free-form natural language prompts. 

\end{abstract}

\begin{figure}
    \centering
    \includegraphics[width=\linewidth]{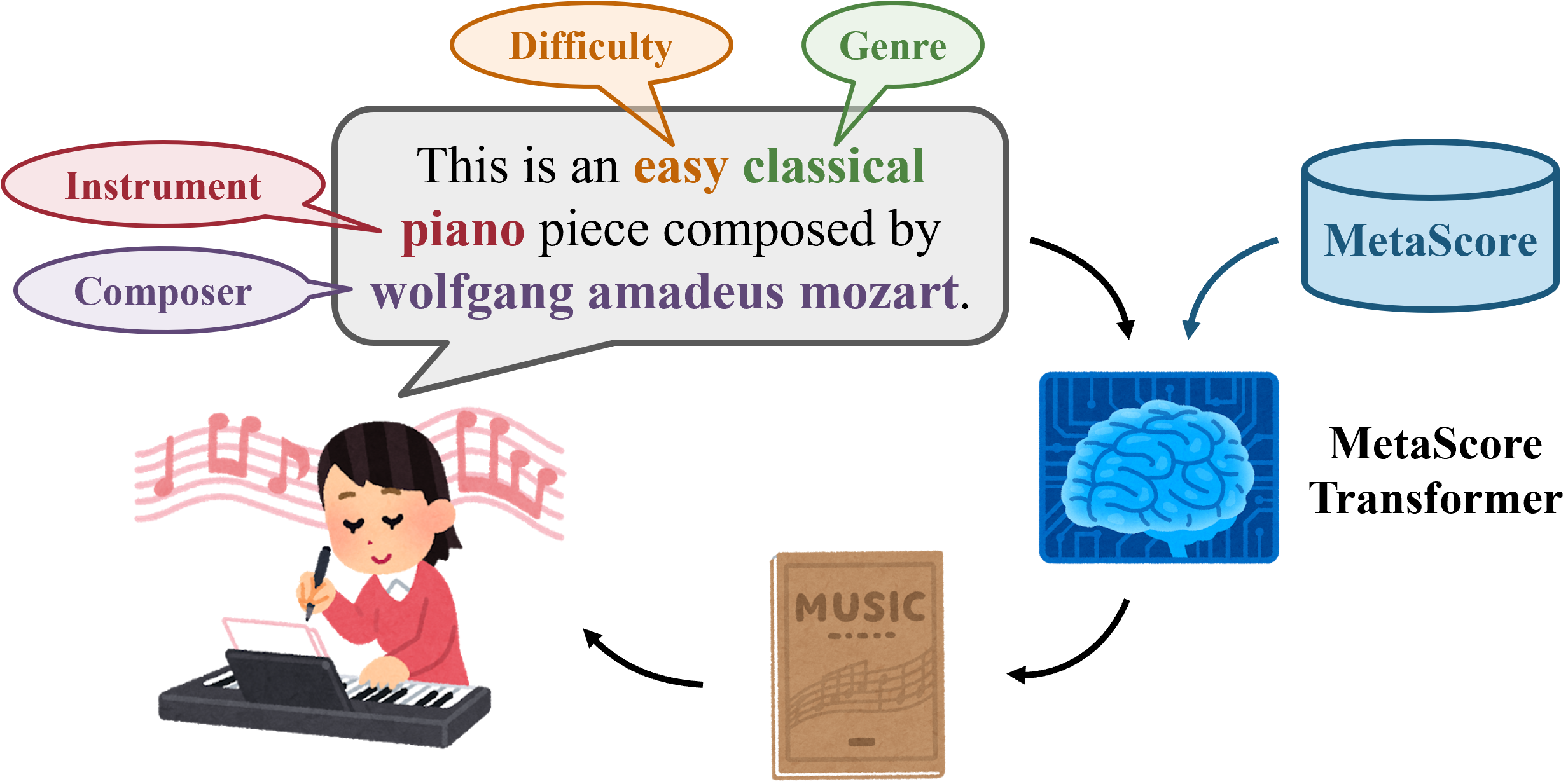}
    \caption{Leveraging the LLM-enhanced MetaScore dataset, our proposed MetaScore Transformer (MST) model generates symbolic music using natural language prompts with difficulty, genre, instrument and composer controls. The symbolic music outputs allow the user to further edit and complete the composition.}
    \label{fig:overview}
\end{figure}

\section{Introduction}\label{sec:introduction}

Recent work has been investigating the potential of conditional music generation with state-of-the-art machine learning
models. 
In particular, we have seen major progress in audio-domain controllable music generation \cite{MusicControlNet,DITTO}, largely thanks to the vast amount of text-audio pairs for training. Unlike audio-domain music generation, symbolic music generation systems generate music in editable formats that can be further completed by the users, making it easier for musicians to integrate such systems into their creative workflow. However, symbolic-domain controllable music generation has been hindered by the lack of a large, public symbolic music dataset with rich metadata. In this paper, we intend to build a natural language based symbolic music generation system with our new public dataset \textit{MetaScore}. MetaScore contains 963K musical scores paired with rich metadata collected from the MuseScore forum\footnote{\url{https://musescore.com/}\label{musescore}}
as well as extensive metadata such as genre, composer, complexity, time signature, key signature, tempo and user interaction statistics (e.g., number of views, likes and comments).\footnote{We note a concurrent work, MidiCaps \cite{midicap}, which contains 168K MIDI files annotated with genre, emotion tags, and LLM-generated captions. However, it does not include composer names, complexity levels, user statistics, and free-form user-annotated tags (see \cref{tab:commondataset}).}
In order to approach text to music generation, we further enhance the MetaScore dataset by completing missing genre metadata using a machine learning-based genre tagging algorithm and we leverage large language models to convert the metadata into natural language captions.

\begin{table*}
    \small
    \centering
    \begin{tabular}{lccrccccccccc}
        \toprule
        &&&&\multirow{2}{*}[-5pt]{\shortstack{Multi-\\track}} &\multicolumn{6}{c}{Metadata}\\
        \cmidrule(lr){6-11}
                                    &Format &Genre &Samples &       &{\footnotesize Genre} &{\footnotesize Composer} &{\footnotesize Emotion} &{\footnotesize Complexity} &{\footnotesize UIS\tabfn[3]} &{\footnotesize Caption}\\
        \midrule
        LMD \cite{LMD:02}           &MIDI   &Misc. &176,581 &\yes       &\no\tabfn[2] &\no\tabfn[2] &\no &\no &\no &\no\\ 
        MetaMIDI \cite{mmd}            &MIDI   &Misc. &437K &\yes       &\yes  &\yes &\no &\no &\no &\no\\
        WikiMusicText \cite{clamp}       &ABC    &Misc. &1,010\tabfn[1] &\no        &\yes &\yes &\no &\no &\no &\yes\\
        EMOPIA \cite{empoia:02}              &MIDI   &Pop   &1,087   &\no        &\yes &\no &\yes &\no &\no &\no\\
        MAESTRO \cite{maestro}             &MIDI   &Classical &1,276 &\no    &\yes  &\yes &\no &\no &\no &\no\\ 
        MidiCaps \cite{midicap} & MIDI & Misc. & 168K & \yes & \yes & \no & \yes & \no & \no & \yes\tabfn[4] \\
        \cmidrule(lr){1-11}
        MetaScore               &XML    &Misc. &1.27M &\yes       &\yes &\yes &\no &\yes &\yes &\yes\tabfn[4]\\
        \bottomrule
    \end{tabular}\\[\smallskipamount]
    \footnotesize
    \tabfn[1]Only a small subset of WikiMusicText is publicly available at \url{https://huggingface.co/datasets/sander-wood/wikimusictext}\\
    \tabfn[2]Available through error-prone mapping to Million Song Dataset \cite{LMD:02, msd}\quad
    \tabfn[3]User interaction statistics\quad
    \tabfn[4]LLM-generated (see \cref{sec:llm})
    \vspace{-.5ex}
    \caption{Comparison of commonly used publicly available symbolic music datasets}
    \label{tab:commondataset}
\end{table*}

\begin{table*}[ht]
    \small
    \centering
    \begin{tabular}{lcccccccccc}
        \toprule
        &\multirow{1}{*}[-3pt]{\shortstack{Model\\size}} &\multirow{1}{*}[-3pt]{\shortstack{Public\\training data}} &\multirow{1}{*}[-4pt]{\shortstack{Open\\source}} &\multirow{1}{*}[-3pt]{\shortstack{Supports\\drums}} &\multirow{1}{*}[-3pt]{\shortstack{Supports free\\text prompts}} &\multicolumn{4}{c}{Controls}\\
        \cmidrule(lr){7-10}
        &&&&& &Instrument &Genre &Composer &Complexity\\
        \midrule
        FIGARO \cite{figaro:02}     &88.30M &\yes &\yes &\no &\no &\checkmark &\no &\no &\no\\
        MuseCoco \cite{MuseCoCo:03} &203M &\no &\yes &\yes &\no &\yes &\yes &\yes &\no\\ 
        BART-based \cite{bart:02}   &139M &\yes &\yes &\no &\yes &\yes &\yes &\no &\no\\
        \cmidrule(lr){1-10}
        MST-Tags            &87.36M &\yes &\yes &\yes &\no &\yes &\yes &\yes &\yes\\
        MST-Text             &87.44M &\yes &\yes &\yes &\yes &\yes\tabfn &\yes\tabfn &\yes\tabfn &\yes\tabfn\\
        \bottomrule
    \end{tabular}\\
    [\smallskipamount]
    \footnotesize\tabfn These can be achieved by free-form text prompts.
    \vspace{-.5ex}
    \caption{Comparison of controllable music generation systems.}
    \label{tab:Implementation}
\end{table*}

Enabled by the metadata provided in MetaScore, we explore text-conditioned music generation with controllable attributes such as instrument, genre, composer, complexity, and other free-form musical descriptors expressed in natural language. With the LLM-enhanced dataset, we train a
transformer-based text-to-music model using a pretrained large language model to encode the input text prompts (see \cref{fig:overview}). In addition, we train a transformer-based tags-to-music model by prepending the input tags to our proposed music representation. Leveraging the LLM-generated captions for training, the proposed text-to-music model achieves competitive performance against the tag-based model while offering a natural language-based interface that allows free-form text inputs.

To evaluate our proposed models, we compare them with an open-source text-to-symbolic music system \cite{bart:02} and a concurrent work \cite{bhandari2024text2midigeneratingsymbolicmusic} in subjective listening studies. In these studies, we demonstrate that our proposed models outperform the baseline model in terms of coherence, arrangement, adherence, and overall quality, while achieving performance comparable to that of the concurrent work \cite{bhandari2024text2midigeneratingsymbolicmusic}.

Our contributions can be summarized as follows:
\begin{itemize}
    \item We present a new publicly available dataset with musical scores paired with rich metadata and LLM-generated natural language captions.
    \item We train two new models for tag- and text-based controllable symbolic music generation that support instrument, genre, composer and complexity controls.
\end{itemize}

The MetaScore dataset, including the musical scores, metadata and the LLM-generated captions, along with source code and pretrained models, will all be made publicly available upon acceptance. Dataset, codebase and audio samples can be found on our  website.\footnote{\url{https://wx83.github.io/MetaScore_Official/}\label{fn:demo}}

\begin{figure*}
    \centering
    \vspace{1ex}
    \includegraphics[width=.95\linewidth]{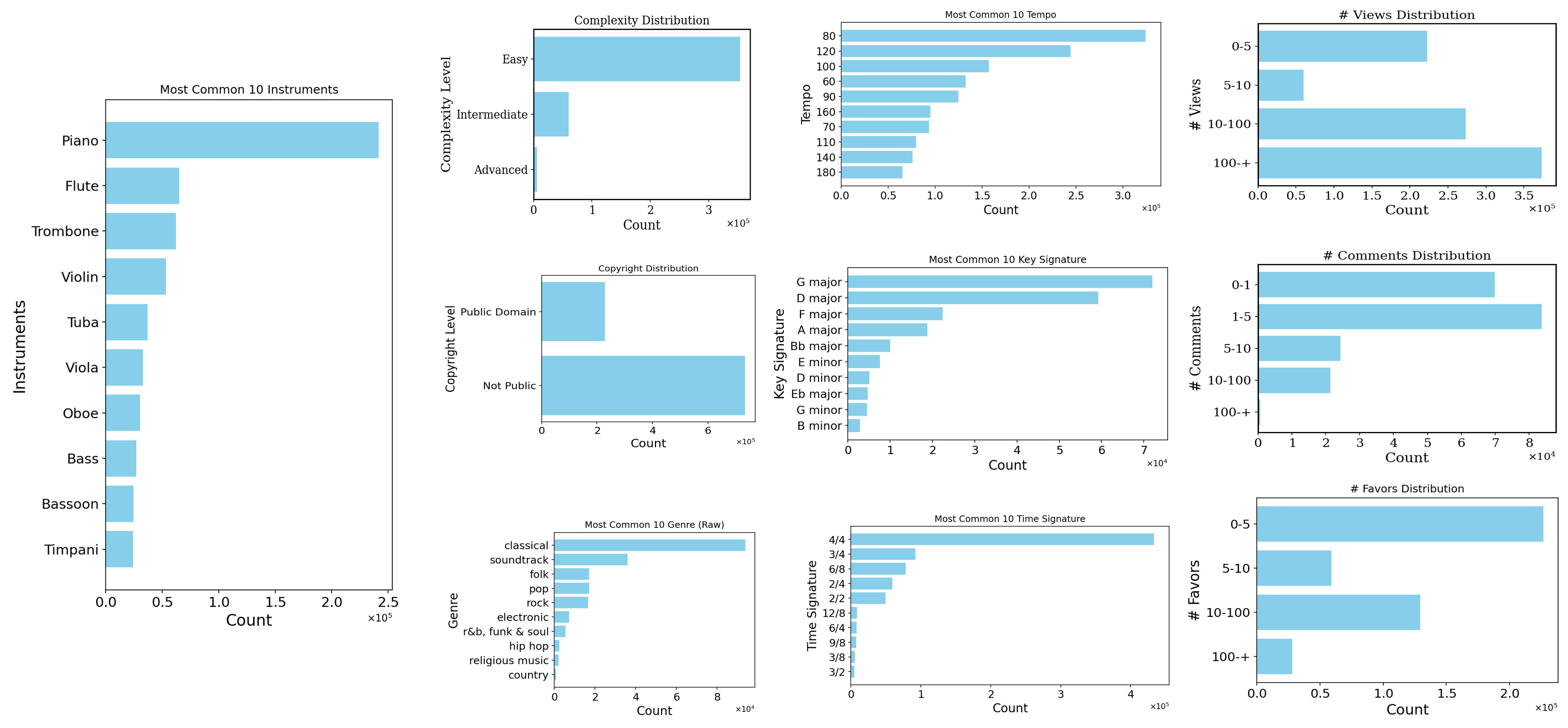} 
    \vspace{-1ex}
    \caption{Statistics of the metadata available in MetaScore-Raw. Note that not all songs come with complete metadata. For each piece, we plot its starting key and starting tempo. To obtain the exact key, we combine user-annotated metadata with MuseScore MusicXML key element (which includes both fifths and mode).}
    \label{fig:distribution}
\end{figure*}

\section{Related Work}

\subsection{Symbolic Music Datasets}
We compare commonly used symbolic music datasets in \cref{tab:commondataset}. 
WikiMusicText\cite{clamp} pairs music with genre, composer, and captions. However, its publicly released version is small, and the musical scores are in ABC notation, which does not support multitrack music natively.
Although MetaMIDI\cite{mmd} comprises around 437K multitrack music pieces in MIDI format, it only includes genre and composer information, lacking 
natural language captions, which are important for training text-to-music generation models.  Although EMOPIA \cite{empoia:02} provides emotion information, it is small and contains only pop music. Although MidiCaps \cite{midicap} contains captions, it does not contain user-annotated free-form tags that are crucial for free-form text-to-music generation. In this work, we present a new, large multitrack and multi-genre symbolic music dataset with rich metadata, including genre, composer, complexity, key signature, time signature, tempo, user interaction statistics, free-form user annotated tags and pseudo captions.

\subsection{Controllable Symbolic Music Generation}
Controllable symbolic music generation include attribute-based music generation, free-form text to music generation and music infilling. We compare our model with existing controllable music generation system in \cref{tab:Implementation}. 
EMOPIA \cite{empoia:02} is designed to generate music that aligns with specific emotional states, defined within the valence-arousal plane. This psychological model categorizes emotions by valence, indicating their positivity or negativity, and arousal, which measures their intensity from calm to excited. FIGARO \cite{figaro:02} can generate samples based on a fine-grained description of the characteristics of the desired music.
MuseCoco \cite{MuseCoCo:03} first classifies a fixed set of predefined musical attributes using multiple classification heads and then employs an attribute-to-music model to generate symbolic music.
Recent work on music infilling \cite{midigpt} can condition generation on attributes including: instrument type, musical style, note density, polyphony level,
and note duration.
In this work, 
we explore free-form text conditioned music generation with LLM-generated natural language music captions. Additionally, we present a tag-conditioned music generation model that can generate music based on four conditions: genre, instrument, complexity and composer.

\section{Dataset}\label{sec:page_size}
\subsection{Dataset Collection}
We scraped our dataset from an open source and free music notation software MuseScore.\cref{musescore}
To evaluate the quality of our dataset, we leverage the rating entry (MetaScore-Raw) as an indicator. This rating, which ranges from 1 to 5 (with 5 being the highest), serves as a structured measure of perceived music quality. To demonstrate that even lower-rated entries can still be suitable for use, we randomly selected 10 samples from three categories: low/missed ratings (below 3 or not rated), mid-range ratings (3–4), and high ratings (above 4). We show  qualitative examples on our demo page.\cref{fn:demo} These examples illustrate the overall usability and diversity of the dataset across different rating levels.

\subsection{Collecting and Preprocessing the Dataset}\label{sec:filter}
We collect 963K songs paired with musical scores and metadata from the MuseScore forum.
We will refer to this original dataset as \textit{MetaScore-Raw}. MetaScore-Raw contains extensive metadata such as genre, composer, complexity, key signature, time signature, tempo, user interaction statistics (e.g., number of views, likes and comments) and free-form user annotated tags. We provide statistics of the metadata in \cref{fig:distribution}, 
and we note that not all songs come with complete metadata.

From the raw MSCZ files, we extract key signature, time signature, tempo and musical instrumentation. We only retain those instruments that are compatible with the General MIDI standard. Regarding composers, we first filter the composer tags to ensure they are formatted as human names and convert them to lowercase. We also standardize the names of well-known musicians to their full names; for instance, ``mozart'' is changed to ``wolfgang amadeus mozart.''

\begin{table}
    \small
    \centering
    \begin{tabular}{l@{~~~~}l@{~~~~}c}
        \toprule
        Type &Dataset &Adherence$\uparrow$\\
        \midrule
        Ground truth genre tags &MetaScore-Genre &3.11 $\pm$ 0.49\\
        Auto-generated genre tags &MetaScore-Plus\tabfn &3.05 $\pm$ 0.54 \\
        LLM-generated captions &MetaScore-Plus & \textbf{3.23 $\pm$ 0.49}\\
        \bottomrule
    \end{tabular}\\[\smallskipamount]
    \footnotesize
    \tabfn We only include songs with auto-generated genre tags here.
    \vspace{-.5ex}
    \caption{Subjective evaluation results on tags/text-music adherence of the dataset are measured on a Likert scale of 1 to 5. We report the mean values and 95\% confidence intervals.}
    \label{tab:taggersubjective}
\end{table}

\subsection{Inferring Missing Genre Tags in MetaScore-Raw}\label{sec:genre_tagger}
While MetaScore-Raw provides rich metadata information, we notice that not all songs come with complete metadata. For example, only 181K (18.8\%) out of 963K songs in MetaScore-Raw contain genre metadata. As genre is one of the most intuitive ways for a user to control the style for a music generation system, we want to complete the genre information for songs without a genre label in MetaScore-Raw. Therefore, we train a genre tagger that is based on the Multitrack Music Transformer (MMT) \cite{MMT:01}, where we remove the causal mask used for autoregressive modeling and append a multi-label classification layer. We select the threshold of the multi-label classification layer for each class based on the F1 score on the validation set.

To infer genre labels for music pieces lacking such tags, we adopt a data-driven approach by training a genre tagger on \textit{MetaScore-Genre}. The genre tagger is based on the Multitrack Music Transformer (MMT) \cite{MMT:01}, where we remove the casual mask used for autoregressive modeling and append a multi-label classification layer. MMT represents a music piece as a sequence of events $x = (x_{1}, \ldots, x_{n})$, where each event $x_{i}$ comprises six attributes: type, beat, position, pitch, duration, and instrument. To create the input sequence, we extract tokens from the start, middle, and end sections of each music pieces, selecting 341 tokens from each to form a concatenated sequence of 1,023 tokens. Instrument information, important for genre identification, is also incorporated as a prefix condition to these sequences.

During training, genres with scarce presence or high ambiguity, such as ``darkwave'' and ``experimental,'' are excluded to avoid noise. We extract notes from the MSCZ files using MusPy \cite{MMT:01}. When we extract notes, we exclude broken files containing negative pitches. Moveover, to enhance generalizability, we include additional 22,000 samples from the LMD dataset \cite{LMD:02}, all tagged with genres.
Due to the rarity or ambiguity of certain genres, we merge specific genre types from these two datasets into 8 classes.\footnote{We define the eight genres as follows: 
``Classical \& traditional'': classical, religious, new age;
``Soundtrack \& stage'': soundtrack, comedy;
``Rock\& Metal'': pop, rock, metal;
``Folk \& country'': folk, country;
``Urban'': hip hop, r\&b, funk\&soul;
``Electronic \& dance'': electronic, disco;
``World'': world music, reggae\&ska;
``Jazz \& blues'': jazz, blues.
}
For the training process,
we allocate 90\% of the samples for training, with 5\% each reserved for validation and testing. We select the threshold of multi-label classification layer for each class based on the performance on the validation set.

To evaluate the performance of genre tagging, we first compute the precision, recall and F1-score on the test set, where we achieve a micro-averaged precision of 61.94, recall of 63.03, and F1 score of 62.48. In addition, we conduct a subjective listening test to compare the quality of the auto-generated genre tags with the user-annotated tags in MetaScore-Raw. The 22 participants are instructed to answer the following question in a Likert scale of 1 to 5: \textit{``How well do you think this piece of music aligns with the following genre?''}. From \cref{tab:taggersubjective}, 
we can see that the auto-generated genre tags in MetaScore-Plus achieves a lower tags-music adherence compared to the ground truth tags in MetaScore-Genre(defined in \cref{sec:dataset_version}), but the difference did not reach statistical significance in our setup.


\subsection{Generating Pseudo Captions using LLMs}\label{sec:llm}
To enable text-based downstream tasks (e.g., music captioning and text-to-music generation), we leverage large language models to convert the metadata into natural language captions. We follow LP-MusicCaps \cite{LPCAP:02} and CLAP \cite{CLAP:02} and adopt an in-context learning-based approach \cite{incontext} using a pretrained large language model.
We form the input prompt string by combining genre, composer, complexity, time signature, key signature, tempo, and free-form user-specified tags.\footnote{We have an old version of LLM-generated captions from tags. MST-Text was trained on an old version of LLM-generated captions in which captions are generated from genre, instrument, complexity, copyright, and free-form descriptors with in-context learning.}
As shown on the demo page,\cref{fn:demo} we provide five examples of input-output pairs to facilitate in-context learning with Bloom\cite{Bloom:02},
where the examples are used to provide guidance
for the LLM to capture the one-to-many mapping between the
input tags and natural language captions. We generate the pseudo captions using the Hugging Face API\cite{scao2022bloom}. We exclude non-English and corrupted captions generated by Bloom\cite{Bloom:02} and truncate the output sequence to a maximum of 32 tokens.

\section{Versions of MetaScore}\label{sec:dataset_version}
We will release the following three versions of MetaScore:
\begin{itemize}
    \item \textit{MetaScore-Raw} (963K): The raw MuseScore files and metadata scraped from the MuseScore forum as well as the corresponding musicxml file for future research.
    \item \textit{Metascore-Genre} (181K): A subset of MuseScore-Raw containing files with user-annotated genres. Additionally, we discard any songs composed by a composer that has less than 100 compositions in MetaScore-Raw. We also provide LLM-generated captions based on information extracted from the metadata in Metascore-Genre. 
    \item \textit{MetaScore-Plus} (963K): MetaScore-Raw where missing genre tags are completed by the trained genre tagger described in \cref{sec:genre_tagger}.
    We also provide LLM-generated captions based on information extracted from the metadata in MetaScore-Plus. 
\end{itemize}
Due to copyright concerns, we will publicly release music scores and metadata that are in the public domain (228K) or licensed with a Creative Commons licenses (46K) from MetaScore-Plus. The rest of the dataset will be provided upon request for research purpose.

\section{Method}\label{sec:typeset_text}

We represent a music piece as a one-dimensional array of integers using an event-based representation adapted from REMI+ \cite{figaro:02} and MMT \cite{MMT:01}. REMI+\cite{figaro:02} represents notes with six consecutive tokens encoding note position, pitch, velocity, duration, instrument and time-signature information. However, it cannot provide control over tags such as genre, composer and complexity. MMT\cite{MMT:01} represents a sequence of six-dimension events, with each event $x_{i}$ encoded as a tuple of variables ($x^\mathit{type}$, $x^\mathit{beat}$, $x^\mathit{position}$, $x^\mathit{pitch}$, $x^\mathit{duration}$, $x^\mathit{instrument}$).
However,
MMT cannot model the interdependencies within these fields for a specific note as it predicts the six fields in parallel.
In this work, we adapt the REMI+ representation \cite{figaro:02} to provide controls over genre, instrument, composer and complexity, while preserving the expressiveness offered by REMI+ \cite{figaro:02}. Similar to MMT \cite{MMT:01}, we decompose note-on events to beat and position to reduce the size of the vocabulary and to help the model learn the rhythmic structure of music. In addition, we exclude the ``tempo'' and ``chord'' events as such information is sometimes unavailable in our dataset. Following REMI+ \cite{figaro:02}, we use \textit{beat}, \textit{position}, \textit{instrument}, \textit{pitch} and \textit{duration} events for representing musical notes for non-drum tracks.
We represent drum notes as \textit{beat}, \textit{position}, \textit{instrument}, \textit{drum\_pitch}.

To enable free-form text controls, 
for each music piece with text, we use a pretrained sentence transformer \cite{sentence:02} (specifically, the ``all-MiniLM-L6-v2'' version \cite{reimers2021allminilm}) to extract the text embedding.
Then we add a linear layer to project the text embedding to the input token embedding space, where the projected text embedding is added to the previous generated token embedding along with the positional encoding.
Then we feed the encoded sequence into a decoder-only linear transformer. We will refer to this model as MetaScore Transformer-Text (MST-Text).

Additionally, we train a tag-conditioned music generation model. To enable tag-based controls, we prepend the input tags to our proposed music representation. We introduce four tag events, including \textit{tag\_genre}, \textit{tag\_composer}, \textit{tag\_complexity} and \textit{tag\_instrument} to specify conditions. Further, we use the standardized composer names to limit the vocabulary size, and we keep only 47 composers that have more than 100 training samples. We use a \textit{tag\_\{missing\_tag\}\_None} event for music pieces that do not contain all four tags. In addition to these data tokens, we have six special structural events: The \textit{start-of-song} event signals the onset of a song, leading into a sequence marked by \textit{start-of-genre}, \textit{start-of-composer}, \textit{start-of-complexity}, \textit{start-of-instrument} events, each followed by their respective tag lists, with \textit{start-of-notes} concluding the tag lists and \textit{end-of-song} indicating the completion of the song.
To facilitate controllability in the model, we prepend these control tokens at the start of the data representation. The control tokens include genre, composer, complexity and instruments. Then we feed the sequence with these prepending tags into a decoder-only linear transformer which capitalizes on the autoregressive nature of the transformer model, enabling the integration of these tokens during the inference process.
We will refer to this model as MetaScore Transformer-Tags (MST-Tags).

\section{Experiments and Results}
\subsection{Baselines}
We compare our model with two text-to-symbolic music generation approaches. The first is a BART-based model \cite{bart:02} trained on a paired text and symbolic dataset using ABC notation, with evaluation presented in \cref{sec:bartobjective} and \cref{sec:bartsub}. The second is a concurrent approach, Text2MIDI \cite{bhandari2024text2midigeneratingsymbolicmusic}, which directly generates MIDI files from natural language prompts, with evaluation presented in \cref{sec:text2midi}.
MuseCoco\cite{MuseCoCo:03} first classifies a fixed set of predefined musical attributes using multiple classification heads and then employs an attribute-to-music model to generate symbolic music. This approach does not support free-form natural language inputs for symbolic music generation. 
The BART-based model \cite{bart:02} leverages pre-trained language models for generating symbolic music in ABC notation. To ensure a fair comparison, we generate music using the BART-based model\cite{bart:02} via the Hugging Face API \cite{reimers2021allminilm} and then convert the ABC outputs to multitrack MIDI using the Melobytes tool \cite{melobytes_abc2midi}.


\begin{table}
    \small
    \centering
    \label{tab:musicgen_objective}
    \begin{tabular}{l@{~~~~}c@{~~~~}c@{~~~~}c}
        \toprule
        &\shortstack{Pitch class\\entropy} &\shortstack{Scale\\consistency} &\shortstack{Groove\\consistency}\\
        \midrule
        MST-Tags-Small  & 2.88 $\pm$ 0.08 & 0.89 $\pm$ 0.02 & \textbf{0.92 $\pm$ 0.01}\\
        MST-Tags & 2.93 $\pm$ 0.07 & 0.89 $\pm$ 0.02 & 0.90 $\pm$ 0.01\\
        \cmidrule(lr){1-4}
        BART-based\cite{bart:02} & 2.54 $\pm$ 0.06 & 0.99 $\pm$ 0.00 & 1.00 $\pm $ 0.00 \\
        MST-Text & \textbf{2.70 $\pm$ 0.06} & \textbf{0.95 $\pm$ 0.01} & \textbf{0.92 $\pm$ 0.01} \\
        \cmidrule(lr){1-4}
        Ground truth &2.67 $\pm$ 0.06 &0.95 $\pm$ 0.01 &0.92 $\pm$ 0.01\\
        \bottomrule
    \end{tabular}
    \caption{Objective evaluation results on music quality with conditions from MST test set. We report the mean values and 95\% confidence intervals.}
    \label{tab:mstonly}
\end{table}

\subsection{Objective Evaluations}\label{sec:bartobjective}

Following \cite{MMT:01,mogren2016crnngan,wu2020jazz}, we assess the quality of generated music using pitch class entropy, scale consistency, and groove consistency, where values closer to the ground truth indicate better performance. To ensure a fair comparison, we randomly sampled 100 conditions from the MST test set and generated corresponding music for evaluation. As reported in \cref{tab:mstonly}, we find that MST-Text most closely matches the ground truth in terms of pitch class entropy and scale consistency, while MST-Tags-Small and MST-Text perform similarly on groove consistency. Additionally, our proposed MST-Text outperforms the BART-based\cite{bart:02} model across all three metrics.

\begin{table*}
    \small
    \centering
    \begin{tabular}{l@{~~~~}c@{~~~~}c@{~~~~}c@{~~~~}c@{~~~~}c@{~~~~}c}
        \toprule
         &Model size &Training samples &Coherence$\uparrow$ &Arrangement$\uparrow$ &Adherence$\uparrow$ &Overall quality$\uparrow$\\
        \midrule
        MST-Tags-Small  & 87.36M &150K &3.87 $\pm$ 0.36 &3.98 $\pm$ 0.38 &\textbf{3.86 $\pm$ 0.38} &3.57 $\pm$ 0.37\\
        MST-Tags  & 87.36M &901K  &\textbf{4.01 $\pm$ 0.37} &\textbf{4.06 $\pm$ 0.39} &3.60 $\pm$ 0.49 &3.66 $\pm$ 0.45\\
        \cmidrule(lr){1-7}
        BART-based\cite{bart:02}    & 139M &283K &3.86 $\pm$ 0.30 &3.63 $\pm$ 0.39 &2.81 $\pm$ 0.50 &3.29 $\pm$ 0.42\\
        MST-Text   & 87.44M &560K &3.93 $\pm$ 0.28 &3.88 $\pm$ 0.33 &3.35 $\pm$ 0.44 &\textbf{3.69 $\pm$ 0.33}\\
        \bottomrule
    \end{tabular}
    \caption{Subjective evaluation results in a Likert scale of 1 to 5}
    \label{tab:musicgen result}
\end{table*}

\begin{table*}
    \small
    \centering
    \begin{tabular}{lcc@{~~~~}c@{~~~~}cc@{~~~~}c@{~~~~}cccccccccccc}
        \toprule
        &\multirow{2}[2]{*}{\shortstack{Model\\size}} &\multicolumn{3}{c}{CLAP Score$\uparrow$} &\multicolumn{3}{c}{Coherence(\%)$\uparrow$} &\multicolumn{3}{c}{Arrangement(\%)$\uparrow$} &\multicolumn{3}{c}{Adherence(\%)$\uparrow$} &\multicolumn{3}{c}{Overall quality(\%)$\uparrow$}\\
        \cmidrule(lr){3-5} \cmidrule(lr){6-8} \cmidrule(lr){9-11} \cmidrule(lr){12-14} \cmidrule(lr){15-17}
        &&M &T &M+T &M &T &M+T &M &T &M+T &M &T &M+T &M &T &M+T \\
        \midrule
        Text2MIDI \cite{bhandari2024text2midigeneratingsymbolicmusic}  &159M &0.23 &0.20 &0.22  &0 &40 &20 &40 &\textbf{50} & 45  &40 &\textbf{80}& \textbf{60} &40 &\textbf{60}& \textbf{50}\\
        MST-Text   & 87.44M  &\textbf{0.36} &0.13 &\textbf{0.24}  &\textbf{100} &\textbf{60} &\textbf{80}  &\textbf{60} &\textbf{50} & \textbf{55} &\textbf{60} &20& 40  &\textbf{60} &40  &\textbf{50}\\
        \bottomrule
    \end{tabular}
    \caption{
Comparison of MST-Text and Text2MIDI\cite{bhandari2024text2midigeneratingsymbolicmusic} on three prompt sets: 1) \textbf{M}: five prompts from our test set, 2) \textbf{T}: five prompts from the Text2MIDI \cite{bhandari2024text2midigeneratingsymbolicmusic} test set, and 3) \textbf{M+T}: the union of these two prompt sets. We report the winning rates in a subjective A/B listening test described in \cref{sec:text2midi}.} 
    \label{tab:textmodelabtest}
\end{table*}

\subsection{Subjective Evaluation}\label{sec:bartsub}

We conduct a subjective test where 22 participants are instructed to evaluate five songs under each scenario.
Out of the 22 participants, 19 people have experience in playing instruments, with two being professional musicians. 
We ask the participants to evaluate the audio samples in terms of coherence, arrangement, adherence and overall quality in a Likert scale of 1 to 5.

We report the subjective evaluation results in \cref{tab:musicgen result}. 
When contrasting MST-Tags-Small with MST-Tags, we observe
that MST-Tags achieves better performance in coherence and arrangement, but we see a decrease in adherence, possibly due to the incorporation of some auto-generated tags. This comparison illustrates the trade-off between employing a smaller, high-quality dataset (MetaScore-Genre) versus a larger yet noisy
dataset (MetaScore-Plus). 
However, comparing the overall quality score of MST-Tags-Small and MST-Tags, we see
that training with a larger dataset leads to an increase in the overall quality of music generation.

For text-conditioned music generation, 
MST-Text outperforms the BART-based\cite{bart:02} approach, in terms of coherence, arrangement, adherence, and overall quality.
We observe that the text-conditioned system MST-Text has a lower adherence against MST-Tags-Small and MST-Tags. 
This implies that text-to-music generation is a more challenging task than tag-to-music generation as a text-to-music generation system needs to learn to interpret the free-form text inputs.

Overall, the tag-conditioned systems, MST-Tags and MST-Tags-Small, demonstrate strong performance across multiple dimensions, including coherence, arrangement, adherence to prompts, and overall music quality. These results highlight the high quality of our constructed dataset. Notably, MST-Text achieves the highest score in overall quality, indicating that our text-conditioned model performs on par with the tag-conditioned variants. This underscores the effectiveness of our approach, which leverages a large language model to generate natural language captions, enabling end-to-end training for high-quality text-to-music generation.

\subsection{Comparison to Text2MIDI}\label{sec:text2midi}

In this section, we compare our proposed MST-Text with a concurrent work Text2MIDI\cite{bhandari2024text2midigeneratingsymbolicmusic} that also supports text-to-symbolic music generation. For a fair comparison, we create three test sets: 1) five text prompts randomly selected from our test set, 2) five text prompts randomly selected from the Text2MIDI test set, and 3) the union of the previous two test sets (i.e., five prompts from each test set).

\vspace{1ex}
\noindent\textbf{Objective Evaluation.}\quad
We compare music quality of our MST-Text and Text2MIDI \cite{bhandari2024text2midigeneratingsymbolicmusic} using pitch class entropy, scale consistency, and groove consistency, and report the results in \cref{tab:obj_joint}. In addition, to assess the alignment between text and symbolic music, we report the average semantic similarity computed with CLAP \cite{CLAP:02}. We find that MST-Text better matches the ground truth in terms of pitch class entropy and scale consistency, while Text2MIDI better matches the ground truth in terms of groove consistency. Additionally, MST-Text shows better alignment performance when prompts are taken from the MST-Text test set and the joint test set.

\begin{table}[ht]
    \small
    \centering
    \begin{tabular}{l@{~~~~}c@{~~~~}c@{~~~~}c}
        \toprule
        &\shortstack{Pitch class\\entropy}  &\shortstack{Scale\\consistency} &\shortstack{Groove\\consistency}\\
        \midrule
        Text2MIDI\cite{bhandari2024text2midigeneratingsymbolicmusic}& 2.44 $\pm$ 0.19 & 0.89 $\pm$ 0.03 & \textbf{0.94 $\pm$ 0.01}  \\
        
        MST-Text & \textbf{2.65 $\pm$ 0.08} & \textbf{0.96 $\pm$ 0.02} & 0.92 $\pm$ 0.01\\
        \cmidrule(lr){1-4}
        Ground truth &2.71 $\pm$ 0.07 &0.96 $\pm$ 0.02 &0.94 $\pm$ 0.01\\
        \bottomrule
    \end{tabular}
    \caption{Objective evaluation results on music quality for the joint test set. We report the mean values and 95\% confidence intervals.}
    \label{tab:obj_joint}
\end{table} 
\vspace{1ex}
\noindent\textbf{Subjective Evaluation.}\quad
We compare our model MST-Text with Text2MIDI \cite{bhandari2024text2midigeneratingsymbolicmusic} via an A/B test on coherence, arrangement, adherence, and overall quality. Eleven participants (9 with musical experience, including one professional) evaluated 5 prompts from our test set and 5 from the Text2MIDI test set, with results summarized in \cref{tab:textmodelabtest}.

Our experiments show that MST-Text produces more coherent results and achieves equal or superior arrangement performance compared to Text2MIDI\cite{bhandari2024text2midigeneratingsymbolicmusic}. However, when it comes to adherence, Text2MIDI outperforms MST-Text on the Text2MIDI test set and the joint test sets, while MST-Text performs better on MST-Text test set. Overall, both models deliver comparable quality.

\section{Conclusion}

In this paper, we have introduced MetaScore, a new publicly available dataset containing rich metadata and LLM-generated captions. We also present a new music generation model that can generate symbolic music from free-form text, allowing controls over instruments, genre, composer, complexity, among
other 
music descriptors.
In addition, the LLM-generated pseudo captions contain information provided in free-form user-annotated tags, which can pose a challenge to systems that adopt a predefined set of tags \cite{MuseCoCo:03}.
Our objective and subjective evaluation results show the effectiveness of the proposed tags-to-music and text-to-music models. Our proposed text-to-music model outperforms a baseline text-to-music model \cite{bart:02} and achieves comparable performance with a concurrent work\cite{bhandari2024text2midigeneratingsymbolicmusic}. In addition, the proposed text-to-symbolic music generation model trained with LLM-generated pseudo captions achieves competitive performance against the proposed tags-to-music model trained using only the ground truth tags.

\section{Ethics Statement}

We note that MetaScore contains many copyrighted contents. This raises concerns of potential misuse of this dataset that can lead to severe copyright infringements. To minimize such risks, we will release those in public domain only and those not in public domain will be shared upon request and only use for research pruporse. Further, music generation systems built upon the MetaScore dataset may infringe the copyright held by the content creators, and thus we must be careful about adopting these systems in commercial applications. However, we would like to point out that when used properly and with caution, a text-to-music generation system can also make a positive impact to society by enabling new opportunities and interfaces for music creation, as demonstrated in \cite{huang2020aisongcontesthumanai}. Given the editable nature of symbolic music, we hope our proposed text-to-symbolic music models will open up new pathways towards human-AI music co-creation.

\bibliography{ISMIRtemplate}

%
%
%
%

\end{document}